\documentclass[twocolumn,nofootinbib,
showpacs,prl,aps,floatfix]{revtex4}

\usepackage{graphicx}

\newcommand {\fexp} [1] {\exp \left( #1 \right)}
\newcommand {\fabsq}[1] {\left| #1 \right|^2}
\newcommand {\fabs}[1] {\left| #1 \right|}

\newcommand {\si} {/\mbox{s}}
\newcommand {\ms} {\, \mbox{ms}}

\newcommand {\cms}{\, \mbox{cm/s}}

\begin{document}
\title{Momentum interferences of a freely expanding
Bose-Einstein condensate in 1D
due to interatomic interaction change}

\author{A. Ruschhaupt}
\email[Email address: ]{a.ruschhaupt@tu-bs.de}
\affiliation{Institut f\"ur Mathematische Physik, TU Braunschweig, Mendelssohnstrasse 3, 38106 Braunschweig, Germany}
\author{A. del Campo}
\email[Email address: ]{qfbdeeca@ehu.es}
\affiliation{Departamento de Qu\'\i mica-F\'\i sica, Universidad del
Pa\'\i s Vasco, Apdo. 644, 48080 Bilbao, Spain}
\author{J. G. Muga}
\email[Email address: ]{jg.muga@ehu.es}
\affiliation{Departamento de Qu\'\i mica-F\'\i sica, Universidad del
Pa\'\i s Vasco, Apdo. 644, 48080 Bilbao, Spain}

\begin{abstract}
A Bose-Einstein condensate may be prepared 
in a highly elongated harmonic trap with negligible  
interatomic interactions using a Feshbach resonance. 
If a strong repulsive interatomic interaction is switched on and
the axial trap is removed to let the condensate evolve freely
in the axial direction,
a time dependent quantum interference pattern takes place in the short time
(Thomas-Fermi) regime, in which the number of peaks of the momentum
distribution increases one by one, whereas the spatial density barely changes. 
\end{abstract}
\maketitle

% ---------------- FIG. 1 BEGINS ----------------
\begin{figure}
\begin{center}
\includegraphics[angle=0,width=0.9\linewidth]{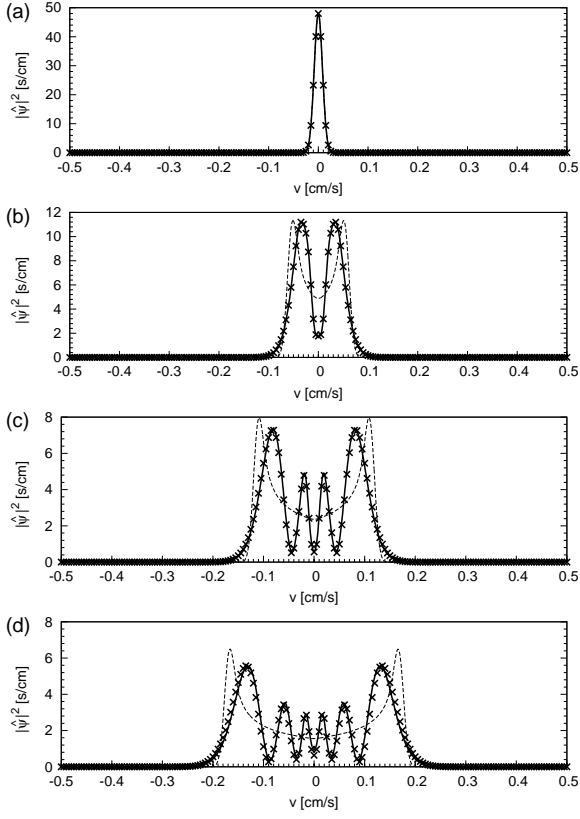}
\end{center}
\caption{\label{fig1}Wave function in momentum space;
exact result $\fabsq{\hat{\psi}(t,v)}$ (lines),
TF approximation $\fabsq{\hat{\psi}_{TF} (t,v)}$ (crosses),
classical result $P (t,v)$ (dashed lines);
(a) $t=0$, (b) $t=0.2\ms$, (c) $t=0.4 \ms$,
(d) $t=0.6\ms$.}
\end{figure}
% ---------------- END FIG. 1 ----------------
%

The dynamics of Bose-Einstein condensates has been much studied, both experimentally and theoretically, for determining the properties of the condensate and its transport behavior in waveguides or free space with potential applications in nonlinear atom optics, atom chips and interferometry. 
Fully free (three dimensional) or dimensionally constrained expansions have in particular been subjected to close scrutiny to obtain, from time series of cloud images and the appropriate theoretical models,   
information on the condensate and/or its confining potentials \cite{Kett99}.   
In reduced dimensions, the expansions have also been examined to characterize and identify different dynamical regimes (the mean-field dominated Thomas-Fermi limit, 
quasi-condensates and the dilute Tonks-Girardeau gas of impenetrable bosons
\cite{OS02,Dettmer01}), or to investigate statistical behavior in a partial release of the 
condensate into a box of finite size \cite{Lew00}.
An effectively one dimensional (1D) Bose gas may be  realized experimentally by a tight confinement of the atomic cloud in two
(radial) dimensions and a weak confinement in the axial direction so that
radial motion is constrained to the ground transversal state. 
Expansions are quite generally described and observed in coordinate space but very interesting phenomena occur in momentum space \cite{Stenger99, Pitaevskii99}.
Also interferences, which show the wave nature of the condensate and may form the basis of metrological applications,
are usually apparent in coordinate space, between independent 
condensates \cite{interf} or as a self-interference \cite{Lew99},
but, as in our present case,    
they may be genuinely momentum space effects,
possibly with an indirect and less obvious spatial manifestation.   
(Other striking example of quantum momentum-space interference phenomenon for an ordinary, non-condensate, one-particle wavefunction colliding with a barrier
has been discussed recently
\cite{mome}.)
The expansions may be manipulated in different ways, e.g. by controlling the
time dependence and shape of the external trapping potentials
or by varying the interatomic interaction
using a magnetically tunable Feshbach resonance \cite{inouye.1998}
or an optically induced Feshbach resonance \cite{theis.2004}.

In this paper we shall take advantage of these control possibilities and 
study a quantum interference effect originating from a change of the
interatomic interaction strength.
The preparation of the condensate may be carried out in a 1D harmonic trap
with negligible interatomic interaction strength such that
the ground state is approximately Gaussian.
Removing the trap under these conditions would preserve the 
momentum distribution, but if the interatomic interaction is
immediately increased as the trap is removed,
the momentum distribution evolution changes
dramatically and in a highly non-classical way:
a time dependent interference occurs consisting of an orderly, one-by-one
increase of the number of peaks in the 
momentum distribution, see Fig. \ref{fig1}, during the early stage of the 1D
expansion, which is described accurately by the Thomas-Fermi (TF) model. In the same
period of time of the snapshots shown, the spatial profile has barely evolved
from the initial profile. Classical mechanics only
explains the global broadening of the momentum distribution due to the release 
of (mean field) interatomic potential energy, notice the motion of the outer
peaks in Fig. \ref{fig1}, but not the oscillatory
pattern, which will require a quantum interference analysis.  
A way to make the momentum interference visible in coordinate
space is to switch off the interatomic interaction again
so that the subsequent free flight maps the momentum peaks into
spatial ones.

\begin{figure}
\begin{center}
\includegraphics[angle=0,width=0.95\linewidth]{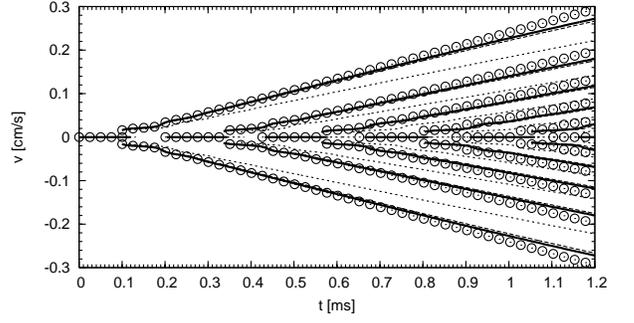}
\end{center}
\caption{\label{fig2}Velocities of the peaks of the wave function
$\fabsq{\hat{\Psi} (t,v)}$:
$g_0 = 0$ (solid lines),
$g_0 = 0.002344\cms$ (dashed lines),
$g_0 = 0.02344\cms$ (dotted lines);
TF approximation $\fabsq{\hat{\psi}_{TF} (t,v)}$ (circles).
In this and similar figures we always apply a threshold level
such that the wave function is assumed to be zero
if $\fabsq{\psi(x)} < 0.001 * \max_{x'}\fabsq{\psi (x')}$.
}
\end{figure}

Let us now discuss the details.          
Assume that an effectively 1D Bose-Einstein condensate
is prepared in a harmonic trap. The condensate wave function is the
ground state of the 1D (stationary) Gross-Pitaevskii equation. 
We shall assume first that the initial interatomic
interaction is zero (this will be relaxed later on).
Then the Gross-Pitaevskii equation
becomes a linear  Schr\"odinger equation so that 
the ground state condensate wave function $\psi_0 (x)$ is a Gaussian, namely
\begin{eqnarray}
\psi_0 (x) &=& \sqrt[4]{\frac{m \omega_x}{\pi\hbar}}
\fexp{-\frac{m\omega_x}{2\hbar} x^2}, 
\label{psi0}
\end{eqnarray}
where $\omega_x$ is the axial (angular) frequency.  
For the rest of the paper we choose $m=\mbox{mass(Na)}$ and
$\omega_x = 5\si$.
At time $t=0$ the confining trap is switched off and the interatomic
interaction is switched on immediately.
(A more realistic switching procedure needing a finite time will be discussed
below.) The time evolution is now described by the 
time-dependent Gross-Pitaevskii equation,
\begin{equation}
i \hbar \frac{\partial\psi(t,x)}{\partial t} 
=- \frac{\hbar^2}{2m}
\frac{\partial^2\psi(t,x)}{\partial x^2}  \nonumber\\
+\, \frac{\hbar}{2} g_1 \fabsq{\psi (t,x)} \psi(t,x),
\label{eqpsi}
\end{equation}
where $g_1$ is the effective 1D coupling parameter. 
We are examining short times $t < t_c := 1/(200\,\omega_x) = 1\ms$
(the factor $1/200$ is arbitrary, it should be true
that $t \ll 1/\omega_x$) for which the absolute square of
the Gaussian in Eq. (\ref{psi0}) under the free evolution
(i.e. Eq. (\ref{eqpsi}) with
$g_1 = 0$) is nearly not changing in coordinate space.
Even with $g_1 = 234.4 \cms$ and for times $t < 1 \ms$,
$\fabsq{\psi(t,x)}$ is nearly not changing.
In contrast, the momentum space distribution changes substantially:
more and more peaks are created as time increases,
see Fig. \ref{fig1}, with $\hat{\psi} (t,v) = \sqrt{\frac{m}{2\pi\hbar}}
\int dx\, \psi (t,x) \fexp{-i \frac{vm}{\hbar} x}$.  
The peak creation can be also
seen in Fig. \ref{fig2} (solid lines), where the velocities of the
peaks versus time are plotted forming a characteristic
structure where bifurcations and creation of a new central peak follow 
each other.  
%
%
% ---------------- FIG. 2 BEGINS ----------------
% ---------------- END FIG. 2 ----------------
%

To understand this effect, we shall approximate
Eq. (\ref{eqpsi}).
If we neglect the kinetic energy (TF regime) we get
\begin{equation}
i \hbar \frac{\partial}{\partial t} \psi_{TF} (t,x)
= \frac{\hbar}{2} g_1 \fabsq{\psi_{TF} (t,x)} \psi_{TF}(t,x).
\label{TF}
\end{equation}
The solution is
$\psi_{TF} (t,x)=\psi_0 (x) \fexp{-\frac{i}{2} t \, g_1 \fabsq{\psi_0(x)}}
$
or, in momentum space,
\begin{eqnarray}
&&\hat{\psi}_{TF} (t,v)
=\sqrt{\frac{m}{2\pi\hbar}}\sqrt[4]{\frac{m \omega_x}{\pi \hbar}}
\int dx\, \exp\bigg[\frac{m\omega_x}{2\hbar} x^2
\nonumber\\
&&- \frac{i}{2} g_1 t \sqrt{\frac{m\omega_x}{\pi\hbar}}
\fexp{-\frac{m\omega_x}{\pi\hbar} x^2} - i \frac{vm}{\hbar} x\bigg].
\label{psiv}
\end{eqnarray}
Note that in the TF regime the spatial density remains unchanged. 
The results for Eq. (\ref{psiv}) are also plotted in Figs. \ref{fig1}
and \ref{fig2}.
There is a good agreement between the exact result and TF, and 
both show the same interference behavior.
It is clear that the nonlinearity is playing a key role in the effect.  

We may compare the quantum dynamics with a similar classical one.
Let us assume an ensemble of classical particles where the probability density
of initial
positions and velocities is given by 
$\rho_0 (x, v) = \fabsq{\psi_0 (x)}\fabsq{\hat{\psi}_0 (v)}$,
with $\psi_0 (v)$ given by Eq. (\ref{psi0}).
The probability density is then evolved with the classical 
Liouville equation 
with the Hamiltonian
$H = \frac{\hbar}{2}g_1 \int dv' \rho_t (x,v')$, without
kinetic term in analogy to the quantum Hamiltonian
of the TF regime, see Eq. (\ref{TF}).
The solution 
is given
by $\rho_t (x,v) = \rho_0 (x, v + \alpha t)$ with
$\alpha = \frac{\hbar}{2} \frac{g_1}{m} \frac{\partial}{\partial x}
\int dv\, \rho_0(x,v)$.
$P (t, v) := \int dx\, \rho_t (x,v)$ is also plotted in Fig. \ref{fig1}. 
The classical picture provides, approximately, the outer peaks moving outwards with time because of the release of potential energy, but not the central ones, which are therefore associated with quantum interference. 
Let us examine this interference.

\begin{figure}[t]
\begin{center}
\includegraphics[angle=0,width=0.95\linewidth]{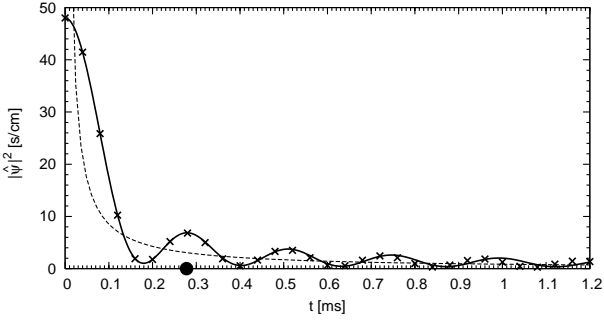}
\end{center}
\caption{\label{fig3} Wave function for $v=0$:
exact result $\fabsq{\hat{\Psi} (t,0)}$ (lines),
TF approximation $\fabsq{\hat{\Psi}_{TF} (t,0)}$ (crosses),
stationary-phase approximation $\fabsq{\hat{\Psi}_s (t,0)}$ (dashed line);
the filled circle marks $t_0$.}
\end{figure}

Defining dimensionless quantities
$\zeta = \sqrt{{m\omega_x}/{\hbar}} \, x$,
$\tau = \sqrt{{m\omega_x}/{\hbar}} \, g_1 t$, and
$\kappa = \sqrt{{m}/{\hbar \omega_x}} \, v$,
we write Eq. (\ref{psiv}) as
\begin{eqnarray}
\lefteqn{\hat{\psi}_{TF} (\tau, \kappa) =
\frac{1}{\sqrt{2\pi\omega_x}}\sqrt[4]{\frac{m\omega_x}{\pi\hbar}}} & &
\nonumber\\
&\times & 
\int d\zeta\, \exp \left[-\frac{\zeta^2}{2} - i \tau
\left(\frac{1}{2\sqrt{\pi}}  e^{-\zeta^2} + \frac{\kappa}{\tau}\zeta\right)
\right].
\label{approx}
\end{eqnarray}
An important characteristic time is the value $\tau_0$ when the second
maximum of $\fabsq{\hat{\psi}_{TF} (\tau_0, \kappa \equiv 0)}$
in the TF
approximation appears (the first maximum is at $\tau=0$),
see Fig. \ref{fig3}. This sets the scale of the oscillations
and the value is found numerically, 
\begin{eqnarray*}
\tau_0 \approx 27.703 \quad \Rightarrow \quad
t_0 \approx \frac{27.703}{g_1}\sqrt{\frac{\hbar}{m\omega_x}}.
\end{eqnarray*}
This time $t_0$ should be much smaller than the critical time $t_c$
when the kinetic energy starts to influence. Therefore we get a
necessary condition for $g_1$, 
\begin{eqnarray*}
& &  t_c = \frac{1}{200\,\omega_x} >
 t_0 \approx \frac{27.703}{g_1}\sqrt{\frac{\hbar}{m\omega_x}}\\ 
&\Rightarrow& g_1 > 27.703 \times 200
\sqrt{\frac{\hbar\omega_x}{m}}\approx 65.12 \cms.
\end{eqnarray*}
%
%
% ---------------- FIG. 3 BEGINS ----------------
% ---------------- END FIG. 3 ----------------
%
% ---------------- FIG. 4 BEGINS ----------------
\begin{figure}[t]
\begin{center}
\includegraphics[angle=0,width=0.95\linewidth]{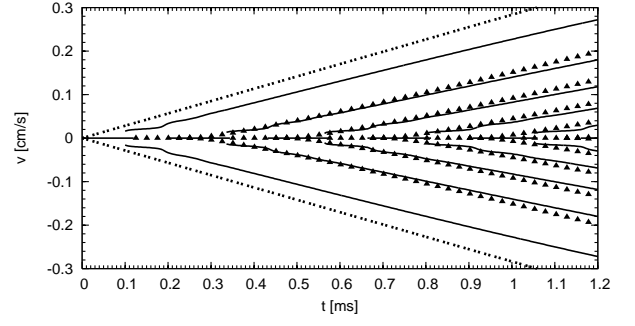}
\end{center}
\caption{\label{fig4} Velocities of the peaks
of the
exact result $\fabsq{\hat{\psi}}$ (lines),
the stationary-phase approximation $\fabsq{\hat{\psi}_s}$ (triangles);
the outer thick-dotted lines mark the critical values $\pm v_c(t)$.}
\end{figure}
% ---------------- END FIG. 4 ----------------
%
%
%
%
We will simplify further Eq. (\ref{approx})
by means of the stationary phase method.
If $\kappa/\tau$ is constant and $\tau \to \infty$ then the main contribution
to the integral (\ref{approx}) comes from the values $\zeta$ which
fulfill
$$
\frac{\partial}{\partial\zeta} \left[
\frac{1}{2\sqrt{\pi}}  e^{-\zeta^2} + \frac{\kappa}{\tau}\zeta
\right]=0 
\Rightarrow -\zeta \frac{e^{-\zeta^2}}{\sqrt{\pi}} + \frac{\kappa}{\tau}=0.
$$
If $0 < \fabs{\kappa/\tau} < \fexp{-1/2}/\sqrt{2\pi}$ there are two solutions
$\zeta_0$, $\zeta_1$ with
$0 < \fabs{\zeta_0} < 1/\sqrt{2} < \fabs{\zeta_1}$.
Physically these are two positions in which the force exerted on the atom is equal 
because the Gaussian profile of the potential changes from concave-down
around the center, to concave-up in the tails. Thus these two positions
contribute to the same momentum and the corresponding amplitudes will interfere
quantum mechanically.   
The stationary phase approximation of Eq. (\ref{approx}) is
\begin{eqnarray}
&&\hat{\psi}_s (\tau, \kappa) = 
\sqrt[4]{\frac{m\omega_x}{\hbar}}
e^{i \frac{\pi}{4}} \frac{1}{\sqrt{\tau\omega_x}}
\label{approx2}\\
&\times\!\!&\!\!\left\{\!
\frac{\exp\left[-i\tau(\frac{e^{-\zeta_0^2}}{2\sqrt{\pi}}+\frac{\kappa}{\tau}\zeta_0)\right]}
{\sqrt{1-2\zeta_0^2}}
- i
\frac{\exp\left[-i\tau(\frac{e^{-\zeta_1^2}}{2\sqrt{\pi}}+\frac{\kappa}{\tau}\zeta_1)\right]}
{\sqrt{2\zeta_1^2-1}}\!
\right\}\!.
\nonumber
\end{eqnarray}
Fig. \ref{fig4} shows also the peaks of $\fabsq{\hat{\psi}_s (\tau, \kappa)}$
resulting from using Eq. (\ref{approx2}) in
the range $0 < \fabs{\kappa/\tau} < \fexp{-1/2}/\sqrt{2\pi}$.
Except for the outer peaks and the $\kappa = 0$ line, Eq. (\ref{approx2}) gives the correct peak behavior.
The opening of the cone of the effect can be approximated by
the definition of a critical $\kappa_c (\tau)$ or $v_c(t)$ to make the interference possible,
\begin{eqnarray*}
\frac{\kappa_c}{\tau} = \frac{\fexp{-1/2}}{\sqrt{2\pi}}
\; \Rightarrow \;
\kappa_c (\tau) = \frac{\fexp{-1/2}}{\sqrt{2\pi}} \, \tau,
\end{eqnarray*}
or
$v_c (t) = {\fexp{-1/2}} \omega_x g_1 t/{\sqrt{2\pi}}$.

\begin{figure}
\begin{center}
\includegraphics[angle=0,width=0.9\linewidth]{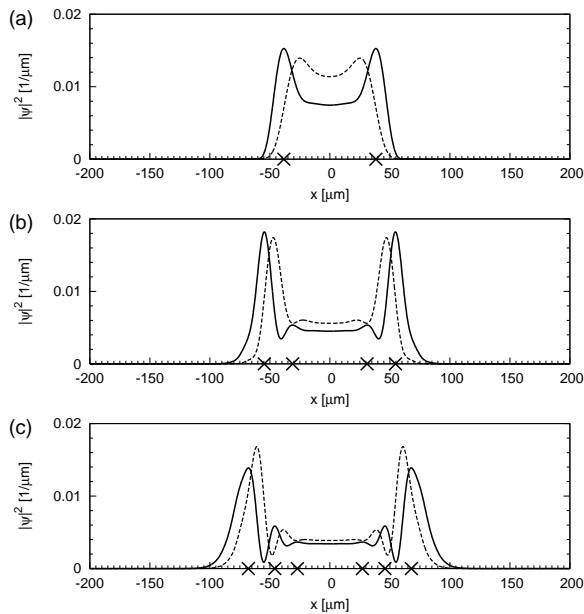}
\end{center}
\caption{\label{fig5}Wave function in coordinate space;
$t = 40 \ms$;
$\Delta t = 0$ (solid lines),
$\Delta t = 0.1\ms$ (dashed lines);
the positions of the maxima are marked by crosses for $\Delta t=0$;
(a) $t_{off}=0.2\ms$,
(b) $t_{off}=0.4\ms$,
(c) $t_{off}=0.6\ms$.}
\end{figure}

The interference pattern in momentum space can be seen in
coordinate space if the interaction is switched
off at a given time $t_{off}$. Then $\fabsq{\hat{\psi}(v,t)}$ is
not changing and therefore the position and number of peaks is not
changing.
On the other hand,
the different peaks separate in coordinate space.
Therefore for different times $t_{off}$ a different
peak pattern will appear in coordinate space (see Fig. \ref{fig5}).
Only the central peak cannot be seen.
%
% ---------------- FIG. 5 BEGINS ----------------
% ---------------- END FIG. 5 ----------------
%

Up to now we have considered an abrupt switching of the interaction,    
but the effect remains for switching times 
$\Delta t$ even of the order of $t_0$. 
For a time dependent coupling parameter given by 
\begin{eqnarray}
g (t)\!=\! g_1\! \times\! \left\{\begin{array}{lcl}
0 &:& t \le 0 \; \mbox{or} \; t \ge t_{off}\\
1 &:& \Delta t \le t \le t_{off}-\Delta t\\
f (t) &:& 0 < t < \Delta t\\
f (t_{off}-t) &:& t_{off}-\Delta t < t < t_{off}
\end{array}\right.,
\end{eqnarray}
with $f (t) = (t/\Delta t)^2 (3 - 2 t/\Delta t)$, 
the result for $\Delta t = 0.1\ms$ is shown 
in Fig. \ref{fig5}:
the effect is not changing qualitatively, only the positions of the
peaks are squeezed.

We shall also examine the stability with respect to 
a non-zero value of the coupling constant used to prepare
the initial state, $g_0$. The initial ground
state is then no longer a Gaussian and can only be calculated
numerically. The effect survives as long as $g_0 \ll
g_1$ but the interference pattern is again squeezed,
see Fig. \ref{fig2}. 

Summarizing we have examined the short-time behavior of the
evolution of a Bose-Einstein condensate in
1D when the interatomic interaction is negligible 
for the preparation in the harmonic trap, and strongly increased 
when the potential trapping is removed.
We have found a quantum interference effect in momentum space originated from
the interatomic interaction change. The momentum distribution
expands due to the
release of mean field energy and the number of peaks increases with time
because of the interference of two positions in coordinate
space contributing to the same momentum or velocity.
The effect is stable in a parameter range and and
could be observed with current technology. 

\begin{acknowledgments}
We are grateful to C. Salomon for encouragement at an early stage of the 
project, and to G. Garc\'\i a-Calder\'on for commenting on the manuscript.
This work has been supported by Ministerio de Educaci\'on y Ciencia
(BFM2003-01003) and UPV-EHU (00039.310-15968/2004).
A.C. acknowledges financial support by the Basque Government (BFI04.479). 
\end{acknowledgments}

\end{document}